\documentclass[prl,twocolumn]{revtex4}

\usepackage{dcolumn}
\usepackage{color}
\usepackage{graphicx}

\begin{document}

\title{{\bf Appearance of room temperature ferromagnetism in Cu-doped TiO$_{2-\delta}$
films.
}}
\author{S. Duhalde and M. F. Vignolo}
\affiliation{Laboratorio de Ablaci\'{o}n L\'aser, Depto$.$ de
F\'{\i}sica, Facultad de Ingenier\'{\i}a, Universidad de Buenos
Aires, Paseo Col\'{o}n 850, 1063 Buenos Aires, Argentina}

\author{C. Chiliotte}
\affiliation{Laboratorio de Bajas Temperaturas, Facultad de
Ciencias Exactas y Naturales, Universidad de Buenos Aires,
Intendente G\"{u}iraldes 2160, 1428 Buenos Aires, Argentina}

\author{C. E. Rodr\'{\i}guez Torres, L. A. Errico, A. F. Cabrera, M.
Renterí\'{\i}a, and F. H. S\'anchez} \affiliation{Departamento de
F\'{\i}sica - Instituto de F\'{\i}sica La Plata (CONICET),
Facultad de Ciencias Exactas, Universidad Nacional de La Plata,
C.C. N$^\circ$67, 1900 La Plata, Argentina}
\author{M. Weissmann}
\affiliation{Departamento de F\'{\i}sica, Comisi\'{o}n Nacional de
Energ\'{\i}a At\'{o}mica, Av$.$ del Libertador 8250, 1429 Buenos
Aires, Argentina}
\date{\today}

\begin{abstract}

We report here the unexpected observation of significant room
temperature ferromagnetism in a semiconductor doped with
nonmagnetic impurities, Cu-doped TiO$_2$ thin films grown by
Pulsed Laser Deposition. The magnetic moment, calculated from the
magnetization curves, resulted surprisingly large, about 1.5
$\mu_B$ per Cu atom. A large magnetic moment was also obtained
from {\it ab initio} calculations , but only if an oxygen vacancy
in the nearest-neighbor shell of Cu was present. This result
suggests that the role of oxygen vacancies is crucial for the
appearance of ferromagnetism. The calculations also predict that
Cu doping favors the formation of oxygen vacancies.
\end{abstract}

\maketitle

Integrating spin functionality into otherwise nonmagnetic
materials has become a highly desirable goal in the last years.
For example, dilute magnetic impurities in semiconductors produce
novel materials appealing for spintronics (see, e.g., refs.
~\onlinecite{ohno98,dietl,fukumura} and references therein). This
is a rapidly developing research area, in which the electron spin
plays an important role in addition to the usual charge degree of
freedom. For their practical applications, ferromagnetic
semiconductors are required to have a high Curie temperature
($T_C$). While most of the dilute magnetic semiconductors (DMS)
like Mn-doped GaAs \cite{ohno96} have a $T_C$ much lower than room
temperature, room-temperature ferromagnetism has been observed in
some Mn-doped compounds such as ZnO
 \cite{sharma}. Recently, Co-doped TiO$_2$ thin films with the anatase structure
were reported to be ferromagnetic even above 400 K with a magnetic
moment of 0.32 $\mu_B$ per Co atom \cite{matsumoto}. These results
have motivated intensive experimental \cite{park,chambers,soo} and
theoretical
 \cite{park2,yang,geng,weng,errico} studies on the
structural and electronic properties of these materials. However,
many questions remain open regarding the underlying microscopic
mechanism of long-range magnetic order. Carrier-induced
interaction between the magnetic atoms was first suggested as the
important ingredient underlying ferromagnetism in III-V based DMS
\cite{munekata}. Subsequent reports have raised concerns about the
initially suggested intrinsic nature of ferromagnetism in these
materials, due to the possibility of ferromagnetic metal
clustering under different growth conditions \cite{sharma,shinde}.
Furthermore, it has been suggested that the strong interaction
between transition metal impurities and oxygen vacancies is
crucial for the existence of high Curie temperature ferromagnetism
\cite{Suryanarayanan,Venkatesanprl,Coey}. We believe our results shed new light
on this problem, suggesting a specific role for the oxygen
vacancies and against the clustering hypothesis.

The original purpose of this work was to find out the role of
oxygen deficiency in the origin and significance of ferromagnetism
in doped TiO$_2$ thin films. Thin films are used because they have
a much greater surface or interface to volume ratio than
single-crystals or polycrystalline ceramics, so the role of
defects may be enhanced. On the other hand, we had previously
found in Fe-doped TiO$_2$ thin films analyzed by X-Ray Absorption
Near-Edge and Extended X-Ray Absorption Fine Structure
Spectroscopy that the magnetic signal increased when the number of
oxygen vacancies around the impurity sites increased
\cite{torres}. In order to produce a similar concentration of
vacancies in TiO$_2$ thin films as in Fe-doped samples but without
the magnetic ions, we studied Cu-doped TiO$_2$ films assuming that
no magnetic signal would come from the dopant. Surprisingly,
significant room temperature magnetic behavior, so strong to give
a magnetization equivalent to 1.5 $\mu_B$/Cu, was found.

\begin{figure}
\includegraphics*[bb= 29mm 165mm 130mm 272mm, viewport=0cm -1cm 16.5cm 11cm, scale=0.5]{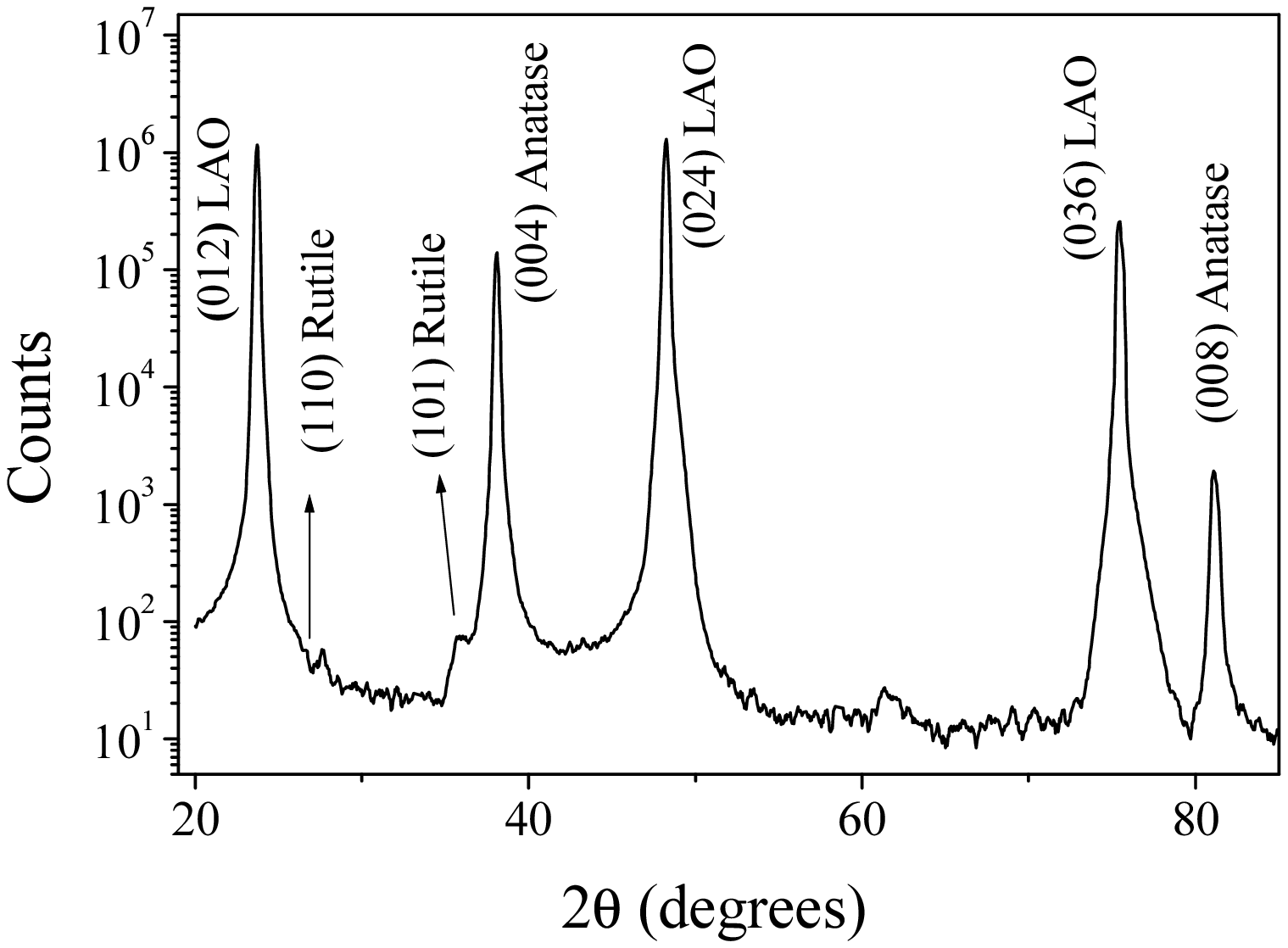}
\caption{\label{fig:xrd} X-Ray Diffraction pattern for Cu-doped
TiO$_{2-\delta}$ film deposited by PLD on a LaAlO$_3$ substrate.}
\end{figure}

Thin films of approximately 10 at.$\%$ Cu-doped TiO$_{2-\delta}$
were deposited on LaAlO$_3$ (001) substrate (LAO) by Pulsed Laser
Deposition (PLD) using a Nd:YAG laser operating at 266 nm. The
TiO$_2$:Cu target was prepared from high purity TiO$_2$ and
metallic Cu powders in stoichiometric quantities. The powders were
mixed for three minutes using a ball-mill, then uniaxially pressed
(200 MPa) into a disk, and finally sintered. The substrate
temperature, laser energy density, oxygen pressure, and pulse
repetition rate were 800 $^\circ$C, 2 J/cm$^2$, 20 Pa, and 10 Hz,
respectively. After deposition the film was cooled down to
room-temperature in 2 hs under 20 Pa oxygen atmosphere. The
composition was determined by Energy Dispersive X-Ray Analysis
(EDX), and no contaminants were found within the accuracy of the
method ($<$ 1 wt. \%). The crystalline structure was studied by
X-Ray Diffraction (XRD). Our films were transparent and strongly
textured, showing only the (001) reflections of the anatase
structure, although rutile reflections were also present but with
much less intensity (Fig. \ref{fig:xrd}). This strong difference
between intensities does not directly reflect the relative amount
of both phases, as our samples consist of a highly oriented
anatase film [only (001) reflections with very large intensities]
and polycrystalline rutile embedded in it [(110) and (101)
reflections of low intensities]. The relative amount of the
anatase and rutile polymorphs depends on the choice of substrate
and on the growth conditions. The lattice mismatch between anatase
TiO$_2$ (001) and LaAlO$_3$ (001) is only 0.26\%, much less than
for rutile, so anatase is favored when deposition is performed on
(001) LAO substrate. However, it has been shown \cite{chambers2}
that a slow growth rate is required to produce anatase single
crystal films of high crystallographic quality, while growing at a
higher rate results in polycrystalline films containing both
anatase and rutile phases. Chambers {\it et al.} \cite{chambers2}
also found that the rutile nanocrystals nucleate within the
evolving anatase film in orientations that maximize the lattice
match between the two phases.

\begin{figure}
\includegraphics*[bb= 36mm 162mm 130mm 272mm, viewport=0cm -2.0cm 16.9cm 10cm, scale=0.56]{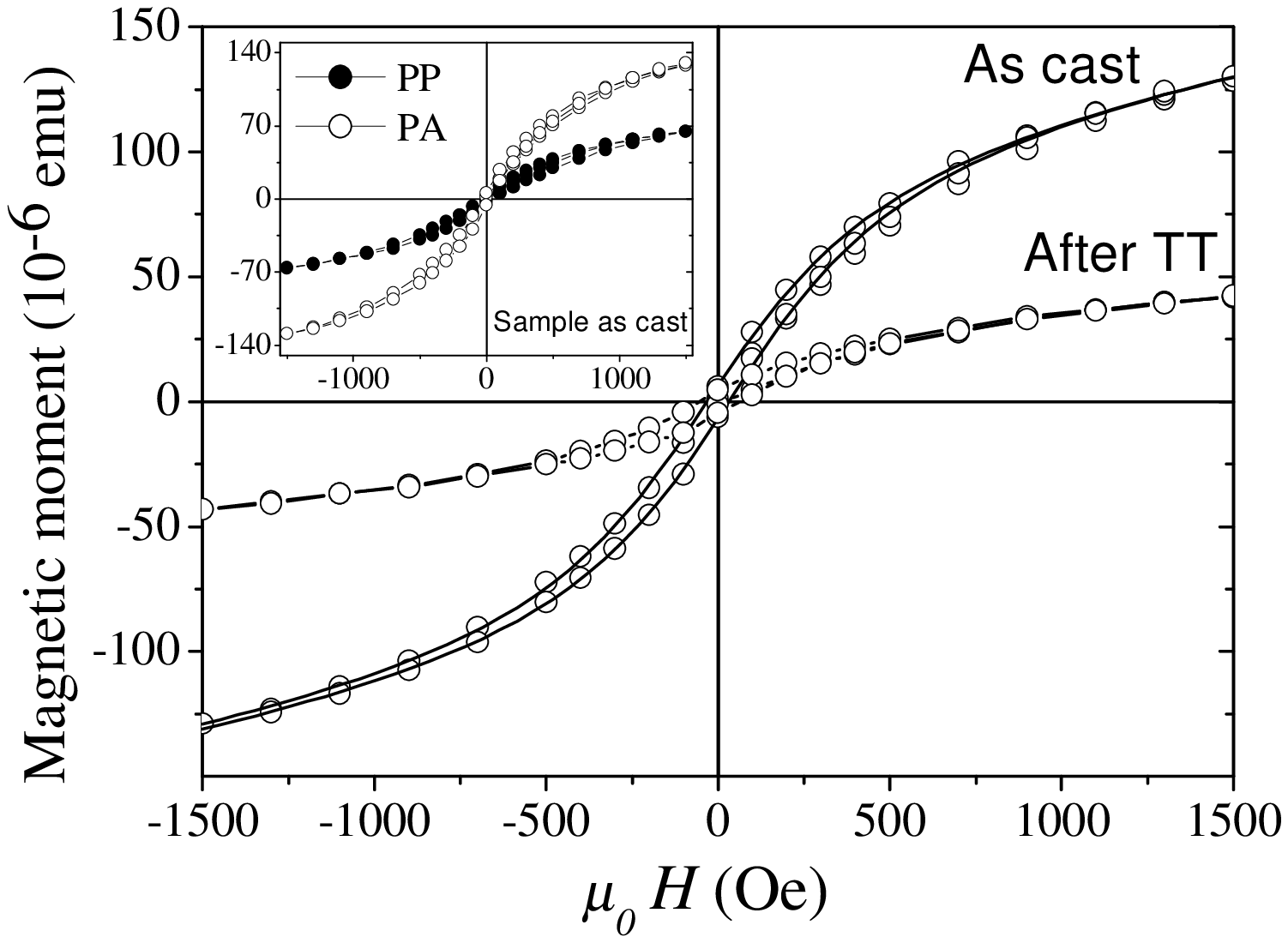}
\caption{\label{fig:mag} Room temperature hysteresis curve of the
Cu-doped TiO$_{2-\delta}$ film (points) in the sample as cast and
after a thermal treatment (TT) at 800 $^\circ$C for 30 min in a
oxygen-rich atmosphere. Solid lines are the best least-squares
fits of Eq.~\ref{1} to the data. The inset compares the hysteresis
loops measured with the external magnetic field applied parallel
(PA) and perpendicular (PP) to the film plane for the as cast
sample.}
\end{figure}

The measurements of magnetization $M$ as a function of the applied
magnetic field $H$ were performed with a commercial Vibrating
Sample Magnetometer Lake Shore 7407 at room temperature, with the
external field applied parallel and perpendicular to the plane of
the film. After subtraction of the diamagnetic contribution due to
LaAlO$_3$ we obtained the result depicted in Fig. \ref{fig:mag},
where significant room temperature magnetization is displayed. In
order to quantify the magnetic parameters (saturation
magnetization $M_s$, intrinsic coercivity $H_c$, and remanent
magnetization $M_r$) we propose the following fitting function for
the demagnetization data:

\begin{equation}\label{1}
M=M_s[(2/\pi)[\arctan((H+H_c)/H_c)\tan(\pi S/2)]] + \chi H,
\end{equation}
where $S=M_r/M_s$. The first term is the usual function used to
represent a ferromagnetic hysteresis curve \cite{stern} and the
second is a linear component representing a possible paramagnetic
contribution. The parameters obtained are:  44$_8$ emu/cm$^3$,
2.5$_4$ emu/cm$^3$, and 35$_3$ Oe for $M_s$, $M_r$, and $H_c$,
respectively. From $M_s$, we estimate an atomic moment of 1.5
$\mu_B$/Cu, assuming 10 at. \% of Cu and a 1000 \AA\ film
thickness.

For the interpretation of this unexpected experimental result we
performed {\it ab initio} calculations on doped TiO$_2$, with and
without oxygen vacancies, using the supercell (SC) method. As a
first approximation we employed a small SC (Ti$_4$O$_8$)
consisting of two unit cells of the rutile structure stacked along
the $c$-axis. One Ti atom was replaced by Cu, giving a rather
large impurity concentration compared with the experimental one.
However, the appearance of a magnetic moment in some cases was
also quite surprising and interesting to relate with the
experiments. The rutile structure was chosen for the calculations
because it gives a more uniform distribution of defects than the
anatase structure when the SC is small. This leads to a more
realistic approximation of the structural distortions induced by
the impurity and the vacancy in the host lattice. The calculations
were performed with the {\it ab initio} full-potential
linearized-augmented-plane-wave method (FP-LAPW) in a scalar
relativistic version, as implemented in the Wien2K code
\cite{wien2k}. We used the local spin density approximation (LSDA)
and the exchange-correlation potential given by Perdew and Wang
\cite{perdew}. The cutoff criterion was R$_{mt}$K$_{MAX}$=7, being
R$_{mt}$ the smallest muffin tin radius and K$_{MAX}$ the largest
wave number of the basis set. The number of $k$-points was
increased until convergence was reached. We also employed,
specially for structural minimizations, the SIESTA code
\cite{siesta} that uses a linear combination of numerical
real-space atomic orbitals as basis set and norm-conserving
pseudopotentials. We proved using Wien2K that the relaxed
structures predicted by SIESTA have lower energies than the
unrelaxed ones and that both methods give similar magnetic
moments.

The SC used for these calculations is tetragonal
($a$=$b$=4.5845$_1$ \AA, $c$'=2$c$=5.9066$_2$ \AA \cite{hill}) and
contains 4 Ti atoms at [(0,0,0); (1/2,1/2,1/4); $R$] and 8 O atoms
at ±[($u$, $u$, 0); (1/2 + $u$, 1/2 - $u$, 1/4); $R$] with
$u$=0.3049$_1$ \cite{hill}. The notation $R$ indicates that the
group of coordinates should be repeated adding (0, 0, 1/2). In
this structure, the Ti atoms are surrounded by a slightly
distorted oxygen octahedron with a rectangular basal plane of four
oxygen atoms with distances to Ti (1.94 \AA) slightly shorter than
those at the vertex (1.98 \AA).

For the case of Ti$_3$CuO$_8$, the system without oxygen
vacancies, we found that the presence of Cu induces a small
distortion of the host lattice, mainly in the six oxygen
nearest-neighbors (O$_{NN}$) of the Cu impurity. The Cu-O$_{NN}$
distances are enlarged from 1.94 \AA\ and 1.98 \AA\ to 1.96 \AA\
and 2.01 \AA, respectively. However, no magnetism was found, which
suggests that the experimentally observed magnetic moment cannot
be due only to the presence of Cu impurities.

In order to analyze the effect of oxygen vacancies in doped
TiO$_2$ we first studied the system Ti$_4$O$_7$, taking into
account the structural distortions produced by the vacancies.
These are much larger than those found in Ti$_3$CuO$_8$, as some
of the Ti atoms are displaced as much as 0.2 \AA. We found that
for this system there are two possible degenerate solutions, one
is non magnetic and the other has a magnetic moment of 1.1 $\mu_B$
per SC, due to polarization of the Ti atoms. It is worth
mentioning that the magnetic phase appears only when structural
relaxation is taken into account. This result anticipates that
oxygen vacancies could play an important role in the magnetic
behavior observed in the TiO$_2$:Cu films. The question that
arises is why the magnetic phase is not observed in bulk rutile
TiO$_2$ without impurities. A possible explanation for this is
that the concentration of intrinsic vacancies formed in
equilibrium is not enough to generate magnetism at the temperature
considered. Calculations with a larger unit cell and only one
vacancy, both in rutile and anatase structures, were studied and a
smaller magnetic moment was obtained for the rutile structure and
none for the anatase structure. So it seems that the appearance of
a magnetic moment may depend on the vacancy concentration and on
the structure. An experimental observation of magnetism in an
undoped nonmagnetic oxide has been recently observed
experimentally and has been attributed to vacancies also
\cite{venkatesan}.

Finally, we studied the oxygen deficient Cu-doped TiO$_2$ systems
(Ti$_3$CuO$_7$). We considered the three possibilities for
removing an oxygen atom from the SC: (i) from the Ti-contained
octahedron, (ii) from the Cu-contained octahedron removing a basal
O atom, and (iii) from the Cu-contained octahedron removing a
vertex O atom. Comparing the total energies of the 3 cases we
found that the oxygen vacancies near Cu are more stable than those
near the Ti site. A similar result was found by Weng {\it et al.}
\cite{weng} for Co in oxygen-deficient TiO$_2$.

The calculations also predicts that the energy required to form an
oxygen vacancy decreases from 10 eV in the undoped system to 4 eV
in the case it is near a Cu impurity. Also, the energy required to
replace a Ti atom by a Cu atom is reduced by 5 eV when there is an
oxygen vacancy in the SC. We therefore predict that Cu doped
systems will have more vacancies than the undoped ones, and that
oxygen vacancies will tend to be close to the impurities. We also
found that the energy required to form two vacancies near a Cu
impurity is larger than that necessary to create one vacancy in
the first shell of neighbors of two different Cu impurities. For
this reason, we discard the formation of vacancy clusters around
Cu atoms.

Concerning the magnetic properties of the system Ti$_3$CuO$_7$, we
found that the magnetic solution has lower energy than the
nonmagnetic one, with a magnetic moment of 1.0 $\mu_B$ per SC, due
to polarization of both Cu and O$_{NN}$ atoms. The value of the
magnetic moment is nearly independent of vacancy location and of
structural relaxation. This result indicates that there is
magnetism in Cu-doped TiO$_2$, but as we have considered only one
SC size (and therefore only one impurity concentration) we cannot
compare its value with the experimental magnetic moment per Cu
atom. To do this we should average over three different
concentrations: that of Cu, that of vacancies, and the relative
amount of rutile and anatase phases. As a first step in this
direction we performed a few calculations with a larger SC
containing 48 atoms. The dimensions of this SC are
$a$'=$b$'=2$a$=2$b$=9.169 \AA, $c$'=2$c$=5.907 \AA. In particular,
for the case of one Cu atom with a neighbor oxygen vacancy in the
SC we obtained a magnetic moment of approximately 1.0 $\mu_B$,
which validates the results obtained for the smaller SC. A similar
result was found for the anatase phase.

\begin{figure}
\includegraphics*[bb=30mm 165mm 190mm 182mm, viewport=0cm -3.5cm 16.5cm 11cm, scale=0.5]{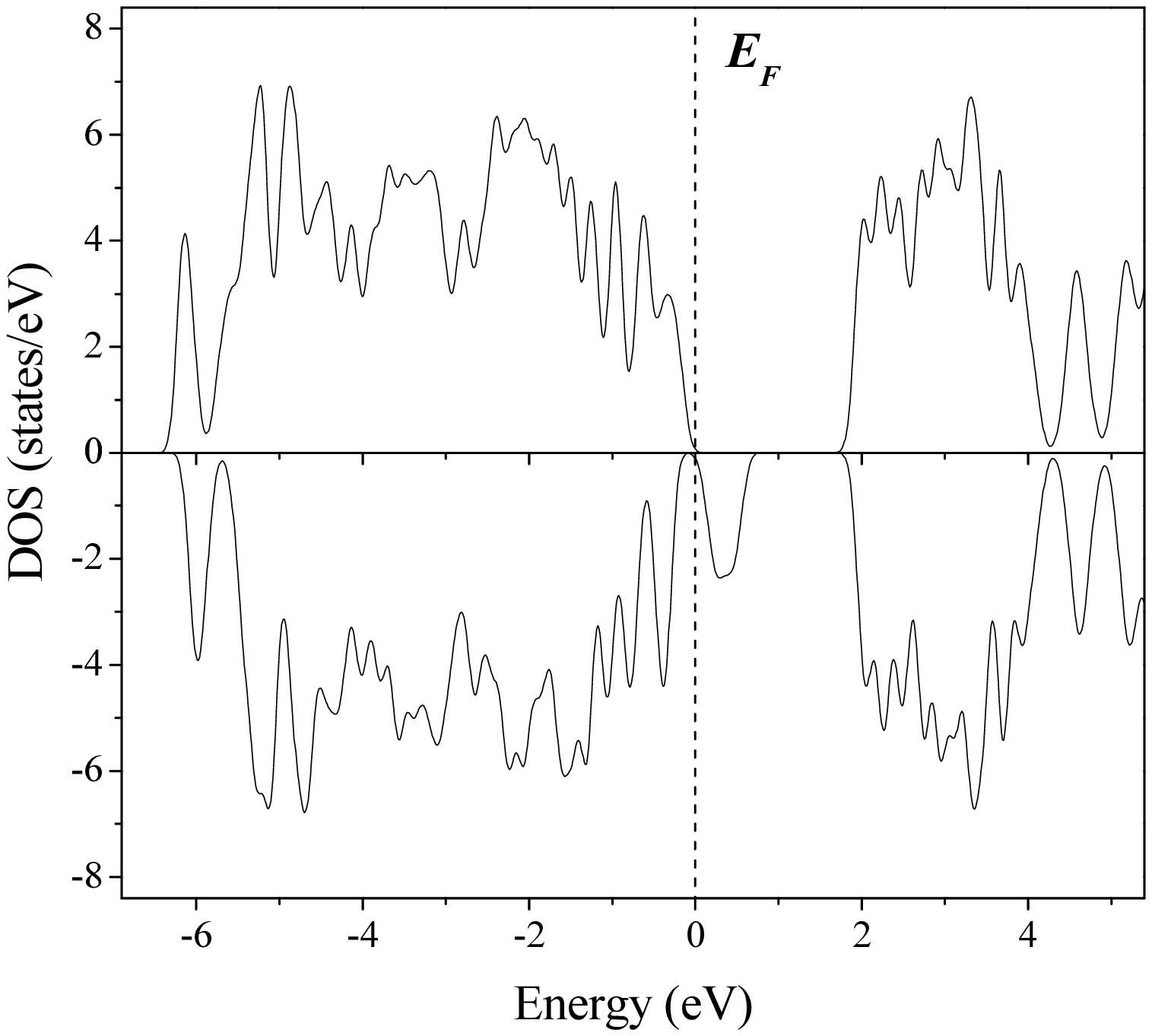}
\caption{\label{fig:dos} Density of states of Ti$_3$CuO$_7$, with
the oxygen vacancy near Cu. Positive values are for spin up and
the negative ones for spin down. Energies are referred to the
Fermi level, $E_F$. The minority spin feature inside the gap but
close to $E_F$ is 50\% of Cu 3d and 50\% of oxygen and Ti
character. }
\end{figure}

Fig. \ref{fig:dos} shows a calculated density of states of the
system Ti$_3$CuO$_7$, which is small gap semiconductor or a
semimetal. However, if the number of vacancies were slightly
larger than that of the Cu impurities, the Fermi level would move
into the minority spin feature close to it. In this case the
system would be magnetic and at the same time half-metallic, with
carriers of only one spin type; this could be very interesting for
spintronics. A similar idea, with excess vacancies responsible for
ferromagnetism was proposed recently \cite{tuan}.

In the previous calculations a ferromagnetic alignment of the Cu
spins was assumed, but we also verified that the pairs Cu-O$_{NN}$
vacancy do not couple antiferromagnetically. This last calculation
was performed for a few particular cases, as the number of
possible distributions of impurities and vacancies is large and so
is the supercell required for that calculation.

In order to verify the crucial role played by the oxygen vacancies
in the origin of ferromagnetism, we performed a "reverse
experiment": in order to reduce the oxygen vacancies, the TiO$_2$
film was heated at 800 $^\circ$C for 30 min in a oxygen-rich
atmosphere. While the XRD spectra show no changes, the
measurements of $M$ as a function of $H$ after the thermal
treatment show a drastic reduction of the magnetic signal (see
Fig. \ref{fig:mag}), in agreement with the  {\it ab initio}
predictions.

In conclusion, we have observed an unexpected significant room
temperature magnetic behavior in Cu-doped TiO$_{2-\delta}$ films,
so strong to give 1.5 $\mu_B$ per Cu atom. This experimental
result indicates that magnetic ions are not essential to obtain
this effect and also that is not due to impurity clustering. {\it
Ab initio} calculations on bulk Cu-doped TiO$_2$, with and without
oxygen vacancies, predict a magnetic moment of 1.0 $\mu_B$ for a
supercell containing one Cu impurity and a neighbor oxygen
vacancy, but no magnetic moment if the oxygen vacancy is absent.
The calculations also predict a lower vacancy formation energy
when there are Cu impurities. Therefore, it seems that oxygen
vacancies play a crucial role for the appearance of magnetism and
that one effect of doping is to increase their number. Both
experimental and theoretical work is now in progress.

This work was partially supported by ANPCyT, PICT98 03-03727,
CONICET, Fund. Antorchas, Argentina, and TWAS, Italy, RGA 97-057.
We thank R. Weht for his assistance with the SIESTA code.

\newpage

\end{document}